# Experimental demonstration of broadband absorption enhancement in partially aperiodic silicon nanohole structures

*Chenxi Lin\*, Luis Javier Martínez, and Michelle L. Povinelli*

Ming Hsieh Department of Electrical Engineering and Center for Energy Nanoscience

University of Southern California, Los Angeles, CA

**Abstract:** We report the design, fabrication, and optical absorption measurement of silicon membranes patterned with partially aperiodic nanohole structures. We demonstrate excellent agreement between measurement and simulations. We optimize a partially aperiodic structure using a random walk algorithm and demonstrate an experimental broadband absorption of 4.9 times that of a periodic array.

Nanophotonic light-trapping structures[1, 2] hold promise to reduce the material cost and improve the efficiency of solar cells. Most of the light-trapping architectures investigated so far consist of perfectly periodic arrays of sub-wavelength features. Recently, however, alternative light-trapping schemes have been proposed based on *aperiodic* photonic patterns. Among these proposals, several have investigated structures with *complete aperiodicity*[3-6] of either random[3, 4, 6] or deterministic[5] nature. Such structures were found to improve solar absorption, with the additional benefit of performance insensitivity to incidence angle and polarization. However, accurate full-field electromagnetic modeling of realistic, three-dimensional, completely aperiodic structures remains unfeasible, severely limiting their design and optimization. Computational work has instead focused on *partially aperiodic* structures[7-15]: such structures are aperiodic on smaller length scales, and periodic on a larger scale (typically on the order of a micron). The large-scale periodicity provides a finite computational cell for calculations, making such structures highly amenable to simulation. Partially aperiodic structures also show strong broadband absorption enhancement relative to their periodic counterparts. Moreover, machine based optimization techniques make it possible to deliberately engineer the partially-aperiodic structure[16], dramatically improving theoretical light-trapping performance[9, 11, 14, 17].

In this paper, we present direct experimental verification of broadband absorption enhancement in partially aperiodic dielectric light-trapping structures. We choose a silicon membrane patterned with nanoholes as our experimental platform. Systematic large-scale numerical simulations show that partially aperiodic nanohole patterns outperform periodic nanohole arrays, given the same hole size and filling ratio. We then use a random-walk algorithm to identify an optimized aperiodic configuration that maximizes solar absorption. Periodic, random, and optimized aperiodic structures were fabricated and optically characterized. Our measurement results confirm the accuracy of our simulations and verify the theoretically predicted increased absorption. We experimentally demonstrate a 4.9 times increase in absorption in the 600nm – 1000 nm range for the optimized aperiodic structure, compared to a periodic array.

Figs. 1(a) and (b) illustrate the geometry of the structures under study. A free-standing silicon nanomembrane with thickness $t$=310 nm is patterned with either periodic (Fig.1 (a)) or partially aperiodic (Fig. 1(b)) nanohole patterns. The periodic pattern is a square lattice of nanoholes with a lattice constant $a$ of 200 nm and a hole diameter $d$ of 120 nm, corresponding to a silicon filling ratio of 71.7%. The partially aperiodic pattern has a disordered pattern of holes within each super cell (dashed lines). The pattern repeats in the $x$ and $y$ directions with periodicity $L$. For fair comparison, the hole size and the silicon filling ratio are the same for the periodic and partially aperiodic structures; for the aperiodic structure, the number of nanoholes within one super cell is $(L/200 \text{ nm})^2$.

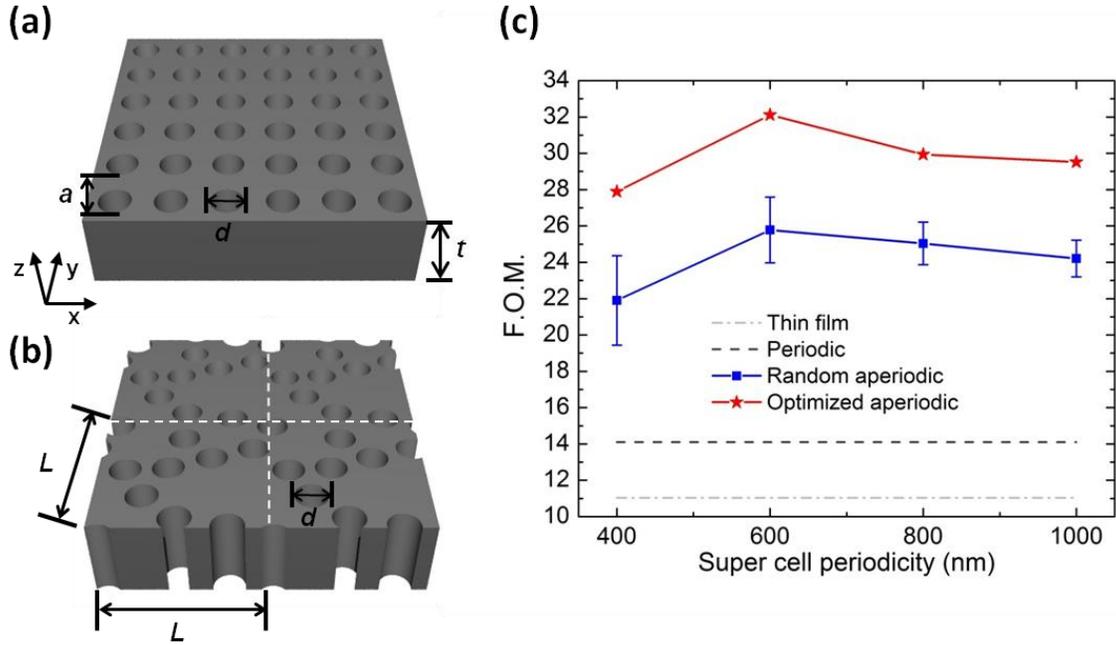

Figure 1 Schematics of the periodic nanohole array (a) and the partially aperiodic nanohole structure (b). (c) Calculated F.O.M. as a function of super cell periodicity in randomly chosen (blue) and optimized (red) partially aperiodic structures. The F.O.M. for the periodic array and an equally-thick thin film are also shown for reference.

We first carried out systematic numerical simulations to predict the effect of partially aperiodicity on broadband absorption. We use the transfer matrix method (TMM)[18] to calculate the absorptance spectra of various structures at normal incidence. Absorptance spectra for two orthogonal polarizations were averaged to obtain the average absorptance for randomly polarized incidence light. We use the tabulated optical constants for crystalline silicon[19] for our simulations.

In order to quantify the broadband optical absorption performance in the solar spectrum, we define a figure of merit (F.O.M.) as follows:

$$F.O.M. = \frac{\int_{\lambda_{min}}^{\lambda_{max}} I(\lambda) A(\lambda) \lambda d\lambda}{\int_{\lambda_{min}}^{\lambda_{max}} I(\lambda) \lambda d\lambda}$$

in which $\lambda$ is the wavelength, $I(\lambda)$ is the ASTM Air Mass 1.5 direct and circumsolar solar irradiance spectrum[20], and $A(\lambda)$ is the average absorptance spectrum of the structure. This F.O.M. represents the ratio between the number of absorbed photons and the total number of incident photons in the incident solar spectrum; $\lambda_{min}$ and $\lambda_{max}$ defines the wavelength range of interest.

Our simulation results are shown in Fig. 1(c). $\lambda_{min}$ and $\lambda_{max}$ were taken to be 400 nm and 1100 nm. The F.O.M. for the periodic nanohole array ($a$=200 nm, $d$=120 nm) and an unpatterned silicon thin film of equal thickness are given for reference. We first investigated *randomly generated* partially aperiodic structures with different super cell periodicities. For each super cell periodicity, a total of 100 configurations were generated by sequentially adding holes at random positions within the super cell. A minimum edge-to-edge nanohole separation of 40 nm was enforced in the generation algorithm due to fabrication constraints. The mean values and standard deviations of the F.O.M. are indicated by blue squares and error bars, respectively. It is clear that random aperiodic structures outperform the periodic structure, regardless of super cell periodicity. The mean F.O.M. for random aperiodic structures depends on super cell periodicity. A super cell periodicity of 600 nm was found to be optimal. This periodicity coincides with the optimal lattice constant for a perfectly periodic nanohole array with the same filling ratio (optimization not shown). For the same super cell periodicity, we observe that different nanohole configurations can have significantly different absorption.

In order to fully exploit partial aperiodicity for absorption enhancement, we carried out random walk optimizations[9] for every super cell periodicity, starting with the same initial periodic configuration described above. The optimization was performed for 100 iterations for each super cell periodicity, subject to the same constraint that the inter-hole distance is larger than 40 nm. The F.O.M.s of optimized structures are shown as red stars. We found that the optimized aperiodic structures have significantly higher F.O.M.s than the mean values of random aperiodic structures, regardless of super cell periodicity. The optimized pattern with the highest F.O.M. also has a super cell periodicity of 600 nm and has a 13.26% higher F.O.M. than the periodic array with a lattice constant of 600 nm and the same filling ratio (F.O.M. = 28.35%).

Drawing on the above analysis, we focus on partially aperiodic structures with a super cell periodicity of 600 nm for our experimental demonstration. The patterns chosen for fabrication are shown in Fig. 2(a – c). Fig. 2(a) illustrates the periodic nanohole array. Fig. 2(b) is the partially aperiodic pattern with a F.O.M. closest to the mean value of our 100 randomly generated configurations. We refer to this configuration as the a*verage aperiodic* pattern. Fig. 2(c) shows the *optimized aperiodic* structure.

We use electron beam lithography combined with inductively coupled plasma reactive ion etching (ICP-RIE) to transfer the nanohole patterns into the device layer of a silicon-on-insulator (SOI) wafer (SOITEC). Every pattern was defined within a circular region with a diameter of 50 µm. After the pattern formation, a 6 mm by 6 mm silicon membrane area was defined by standard UV photolithography and ICP-RIE. The center square part of the silicon membrane with dimensions of 500 µm by 500 µm was aligned to overlap the nanohole patterns, while the rest of the membrane was patterned with access holes to facilitate the wet etching of the buried oxide (BOX) layer. Finally, the silicon membrane was released from the handle silicon wafer in 49% hydrofluoric acid (HF) and wet-transferred[21] to an oxidized silicon wafer. The device area was centered over a perforated window in the silicon wafer to obtain a free-standing membrane.

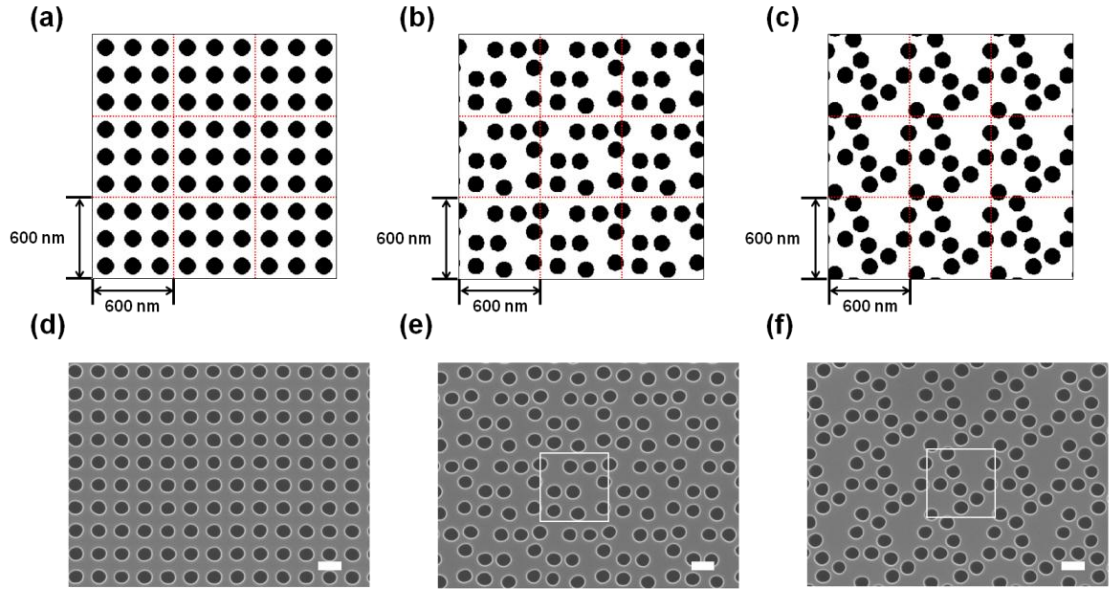

Figure 2 Periodic (a), average aperiodic (b), and optimized aperiodic nanohole patterns. A total of 9 super cells are shown for each configuration. Red dashed lines are the borders between super cells. (d – f) show the corresponding SEM images of the fabricated patterns. The white square denotes a super cell and the scale bars are 200 nm.

Fig. 2(d –f) show scanning electron microscope (SEM) images of fabricated periodic (d), average (e), and optimized (f) partially aperiodic nanohole patterns. Image processing software [22] revealed that the differences in silicon filling ratio between these patterns are below 2%.

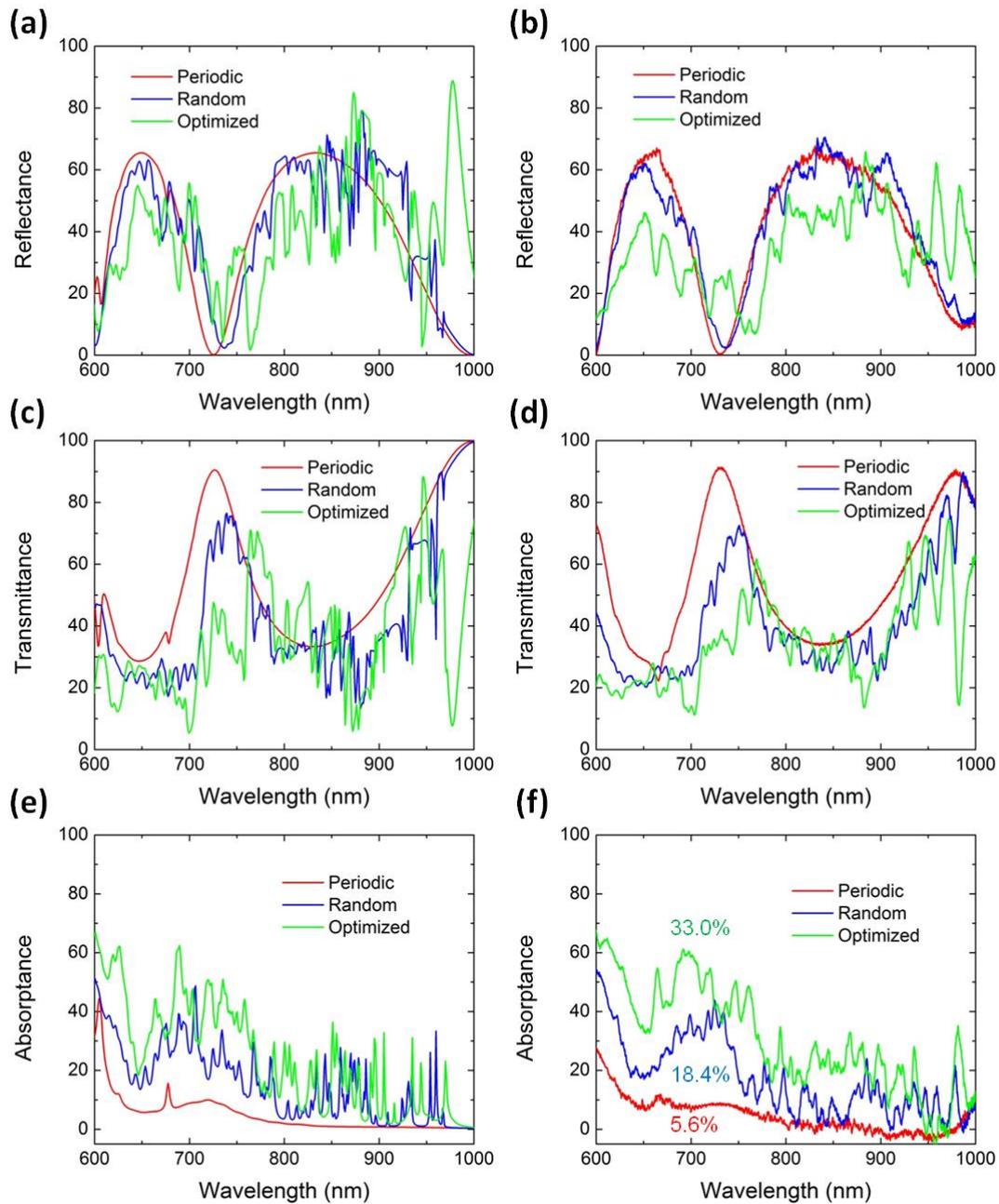

Figure 3 Simulated (left panel) and measured (right panel) reflectance, transmittance, and absorptance spectra of periodic nanohole array (red), average (blue) and optimized (green) partially aperiodic nanohole structures shown in Fig. 2.

We characterized the optical absorption in the fabricated samples in a customized while light spectroscopy setup consisting of a tungsten lamp and a fiber-coupled spectrometer (Ocean Optics USB 4000). The collimated, unpolarized incident light was focused by a microscope objective (10x, N.A.=0.26) to only illuminate the patterned area of interest (circular area with 50 µm diameter). The reflected light from the sample was collected by the same objective, while the transmitted light was collected by an achromatic doublet lens ($f$=30mm, N.A.=0.39). The collected light was focused into multimode fibers connected to the spectrometer for analyzing both the reflectance $R_{exp}(\omega)$ and transmittance $T_{exp}(\omega)$ spectra. The experimental absorptance spectrum $A_{exp}(\omega)$ can then be determined by $A_{exp}(\omega) = 1-R_{exp}(\omega)-T_{exp}(\omega)$. We used a silver mirror as the reflection reference while the transmission reference was air.

We simulate the fabricated structures assuming a silicon membrane thickness of 328 nm and a uniform nanohole diameter of 110nm. The thickness was determined by fitting the measured optical transmittance spectrum of an unpatterned area, and the nanohole size was found to give the best agreement between simulation and measurement. These values are slightly different from those used for our calculations for Fig. 1(c). Nonetheless, simulations with these experimentally determined structural parameters give the same trends as the original simulations. We focus on the spectral data between 600 nm to 1000 nm. For wavelengths below 600 nm, light can be diffracted from the aperiodic structures, leading to incomplete collection by our finite NA lenses. We thus focus on the 600 nm to 1000 nm range, where accurate measurements of optical absorptance can be obtained.

In Fig. 3, we plot the simulated (left column) and measured (right column) spectra. Overall, we observe excellent agreement between simulation and measurement. The experimental spectra tend to be smoother with broader features, which can be attributed to the non-uniformity in the size and shape of nanoholes throughout each pattern[23], as well as the finite angular spread of the incidence beam[24].

We first examine the perfectly periodic array (red lines). We observe characteristic Fabry-Perot fringes in both transmittance and reflectance spectra, with a few guided resonance induced features[25] in the short wavelength range. An experimental value of 5.6% was obtained for the F.O.M. between 600 nm and 1000 nm, based on the data in Fig. 3(f). The average aperiodic structure (blue lines) has a similar reflectance to the periodic structure (Figs. 3(a-b)), but strongly suppressed transmittance (Figs. 3(c-d)), resulting in a broadband absorption enhancement (Figs. 3(e-f)). The experimental value of the F.O.M. is 18.4%. The optimized aperiodic structure (green lines) further reduces the transmittance and improves absorption, offering a F.O.M. of 33.0%. This represents a broadband absorption enhancement 4.79 times higher than a solid film with the same thickness, with only 75.5% of the silicon volume, and 4.9 times higher than a periodic array with the same silicon volume.

In order to qualitatively explain the improvement in optical absorption from periodic, to average, to optimized aperiodic structures, we examine the spatial Fourier spectra. The FFTs of the dielectric structure are shown in Fig. 4(d–f), and the corresponding real space configurations are again shown in Fig. 4(a–c) for reference. The amplitudes of the Fourier components are normalized to the "DC" component of the Fourier transform. The Fourier transform of the periodic square array is also a square lattice, featuring discrete, well-defined peaks with a 200 nm periodicity. Such small length scales can only support the excitation of a few guided resonance modes in the silicon nanohole array for wavelengths above 600 nm, consistent with the spectra shown in Fig. 3(c). As a result, the absorption performance is poor. In contrast, in partially aperiodic configurations, the randomization of the nanohole positions within a super cell generates non-zero spatial frequencies between $(2\pi/200)$ nm$^{-1}$ and $(2\pi/600)$ nm$^{-1}$, corresponding to effective periodicities of 200 nm and 600 nm, respectively. This redistribution of the Fourier components facilitates the coupling between the incident light and in-plane optical resonance modes for wavelength above 600 nm, leading to many more resonance-induced features in the spectra (see Fig. 3(c)) and significantly enhanced broadband optical absorption. Compared to the average aperiodic structure, the optimized structure further suppresses the spatial frequencies corresponding to a 200 nm effective periodicity, while clearly enhancing the spatial frequencies corresponding to a beneficial 600 nm effective periodicity and enabling even better light-trapping.

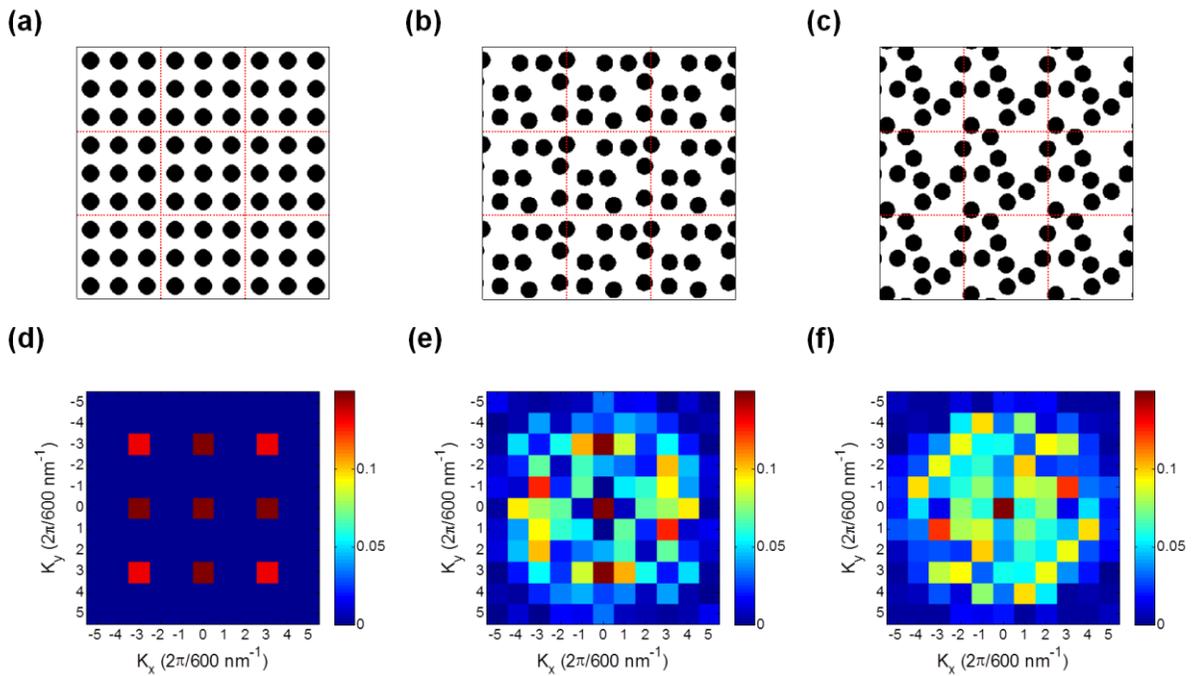

Figure 4 Real space configurations (a – c) and spatial Fourier spectra (d – f) of the periodic nanohole array ((a) and (d)), the average ((b) and (e)) and the optimized ((c) and (f)) partially aperiodic nanohole patterns.

In summary, we demonstrate both numerically and experimentally that silicon membranes patterned with partially aperiodic nanohole structures yield higher broadband optical absorption than those with simple periodic patternings. In particular, *optimized* aperiodic structures perform significantly better (4.9 times in the current study) than periodic ones, though they both contain the same total amount of absorbing material.

Since the random-walk optimization algorithm used in this paper is not guaranteed to find the global optimum structure, our results may be understood as a conservative estimate on performance: further optimization may yield aperiodic structures with even higher broadband absorption. Moreover, we expect that optimization starting from a periodic array with larger lattice constant (600 nm) may yield even more efficient absorbers. We expect that the fundamental optical physics demonstrated here may contribute to the development of highly efficient, thin film photovoltaic devices.

**Acknowledgement**

The authors thank Jing Ma for assistance in microfabrication and Ningfeng Huang, Apoorva Athavale, Bostjan Kaluza, and Milind Tambe for discussions of features of optimal aperiodic structures. Chenxi Lin was supported by the USC Graduate School's Theodore & Wen-Hui Chen Fellowship. Materials and supplies and partial summer salary support for Michelle Povinelli were funded by the Center for Energy Nanoscience, an Energy Frontiers Research Center funded by the U.S. Department of Energy, Office of Science, Office of Basic Energy Sciences, under Award No. DE-SC0001013. Development of the fabrication process for silicon nanohole slabs (Luis Javier Martinez) was funded by the Army Research Office under Award No. 56801-MS-PCS. Computing resources were provided by the USC Center for High Performance Computing and Communication (HPCC).**References**

1. Yu, Z.; Raman, A.; Fan, S. Opt. Express 2010, 18, (S3), A366-A380.

2. Atwater, H. A.; Polman, A. Nat. Mater. 2010, 9, (3), 205-213.

3. Vynck, K.; Burresi, M.; Riboli, F.; Wiersma, D. S. Nat Mater 2012, 11, (12), 1017-1022.

4. Burresi, M.; Pratesi, F.; Vynck, K.; Prasciolu, M.; Tormen, M.; Wiersma, D. S. Optics Express 2013, 21, (S2), A268-A275.

5. Trevino, J.; Forestiere, C.; Di Martino, G.; Yerci, S.; Priolo, F.; Dal Negro, L. Optics Express 2012, 20, (S3), A418-A430.

6. Ferry, V. E.; Verschuuren, M. A.; Lare, M. C. v.; Schropp, R. E. I.; Atwater, H. A.; Polman, A. Nano Lett. 2011, 11, (10), 4239-4245.

7. Chutinan, A.; John, S. Physical Review A (Atomic, Molecular, and Optical Physics) 2008, 78, (2), 023825.

8. Bao, H.; Ruan, X. Opt. Lett. 2010, 35, (20), 3378-3380.

9. Lin, C.; Povinelli, M. L. Opt. Express 2011, 19, (S5), A1148-A1154.

10. Du, Q. G.; Kam, C. H.; Demir, H. V.; Yu, H. Y.; Sun, X. W. Opt. Lett. 2011, 36, (10), 1884-1886.
13